# Securing Fuel for Our Frigid Cosmic Future

Abraham Loeb

Advanced civilizations will likely migrate into rich clusters of galaxies, which host the largest reservoirs of matter bound by gravity against the accelerated cosmic expansion.

______

June 19, 2018

The accelerated expansion of the Universe pushes resources away from us at an ever-growing speed. Once the Universe will age by a factor of ten, all stars outside our Local Group of galaxies will not be accessible to us as they will be receding away faster than light. Is there something we can do to avoid this cosmic fate?

Following the lesson from Aesop's fable[1] "The Ants and the Grasshopper", it would be prudent to collect as much fuel as possible before it is too late, for the purpose of keeping us warm in the frigid cosmic winter that awaits us. In addition, it would be beneficial for us to reside in the company of as many alien civilizations as possible with whom we could share technology, for the same reason that animals feel empowered by congregating in large herds.

After writing a few papers on the gloomy cosmic isolation that is expected in our long-term future[2-5], I received an optimistic e-mail[6] from Freeman Dyson in 2011 where he suggested contemplating a vast "cosmic engineering" project, in which we (in collaboration with any neighboring civilizations, if they exist and cooperate) will concentrate matter from a large-scale region around us to a small enough volume such that it will stay bound by its own gravity and not expand with the rest of the Universe. A similar idea was discussed very recently by Dan Hooper[7], who suggested using the energy output from Sun-like stars to concentrate them across tens of millions of light years. Unfortunately, smaller stars do not produce sufficient power to traverse such distances fast enough. But there are additional limitations to this approach. First, we do not know of any technology that enables moving stars around, and moreover Sun-like stars only shine for about ten billion years (of order the current age of the Universe) and cannot serve as nuclear furnaces that would keep us warm into the very distant future.

Fortunately, mother Nature was kind to us as it spontaneously gave birth to the same massive reservoir of fuel that we would have aspired to collect by artificial means. Primordial density perturbations from the early universe led to the gravitational collapse of regions as large as tens of millions of light years, assembling all the matter in them into clusters of galaxies - each containing the equivalent of a thousand Milky Way galaxies. Therefore, an advanced civilization does not need to embark on a giant construction project as suggested by Dyson and Hooper, but only needs to propel itself towards the nearest galaxy cluster and take advantage of the cluster resources as fuel for its future prosperity. The nearest cluster to us is *Virgo*, whose center is about fifty million light years away. Another massive cluster, *Coma*, is six times farther.

For the above reasons, advanced civilizations throughout the Universe might have migrated towards clusters of galaxies in recent cosmic history, similarly to the movement of ancient civilizations towards rivers or lakes. Once settled in a cluster, a civilization could hop from one star to another and harvest their energy output just like a butterfly hovering over flowers in a hunt for their nectar. The added benefit of naturally-produced clusters is that they contain stars of all masses, much like a cosmic bag that collected everything from its environment. The most common stars weigh a tenth of the mass of the Sun, but are expected to shine for a thousand times longer because they burn their fuel at a slower rate. Hence, they could keep a civilization warm for up to ten trillion years into the future. The nearest examples of dwarf stars in the form of *Proxima Centauri* or *TRAPPIST-1* are known to host habitable Earth-size planets around them, implying that these abundant stars offer attractive parking spots for civilizations which rely on liquid water.

Are there many advanced civilizations out there? Cosmic modesty[8] would suggest an answer in the affirmative, as long as these civilizations do not destroy themselves too quickly[9] (in which case, Darwinian evolution would favor those who are smart enough to sustain longevity). Can we see signs for their long journeys across the cosmic web as their spacecrafts flock collectively on their way towards clusters of galaxies? Probably not, unless they produce powerful beacons of light (for propulsion through lightsails or for communication) that are detectable across the vast cosmological scales[10].

Some lucky civilizations were born in clusters and inherited the resources around them without the need to travel. Could others develop the technology that would enable them to reach the nearest galaxy cluster fast enough? In order to traverse a hundred million light years within the age of the Universe, their spacecrafts need to exceed a percent of the speed of light. This is over a hundred times faster than the speed of all chemical rockets launched thus far by our civilization into space. The *Starshot Initiative*[11] is the first well-funded attempt to develop the technology to propel a spacecraft to a significant fraction of the speed of light. If successful, our civilization could contemplate a future journey to the *Virgo* or *Coma* clusters.

This would be an impressive feat of long-term planning. When looking at photo albums that are billions of years old, our descendants might reminisce on the early millennia that their early technological civilization spent within the Milky Way galaxy. By then, that birth site will be receding away from them at an ever-increasing speed until its image will freeze and fade away for eternity[2].

## ABOUT THE AUTHOR

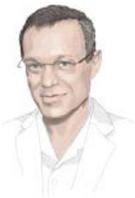

**Abraham Loeb**

Abraham Loeb is chair of the astronomy department at Harvard University, founding director of Harvard's Black Hole Initiative and director of the Institute for Theory and Computation at the Harvard-Smithsonian Center for Astrophysics. He chairs the Board on Physics and Astronomy of the National Academies and the advisory board for the Breakthrough Starshot project.

Credit: Nick Higgins